# Unveiling hole-facilitated amorphisation in pressure-induced phase transformation of silicon

Tong Zhao[1,4,5], Shulin Zhong[2,5], Yuxin Sun[1,5], Defan Wu[1], Chunyi Zhang[3], Rui Shi[2], Hao Chen[1], Zhenyi Ni[1], Xiaodong Pi[1], Xiangyang Ma[1,4✉], Yunhao Lu[2✉], Deren Yang[1,4✉]

[1] State Key Laboratory of Silicon and Advanced Semiconductor Materials and School of Materials Science and Engineering, Zhejiang University, Hangzhou 310027, China

[2] School of Physics, Zhejiang University, Hangzhou 310058, China

[3] Department of Chemistry, Princeton University, Princeton, New Jersey 08544, USA

[4]Shangyu Institute of Semiconductor Materials, Shaoxing 312300, China

[5]These authors contributed equally: Tong Zhao, Shulin Zhong, Yuxin Sun

✉email: mxyoung@zju.edu.cn, luyh@zju.edu.cn, mseyang@zju.edu.cn

Pressure-induced phase transformation of silicon (Si) ubiquitously occurs in most wafering processes such as slicing, grinding and lapping, where high pressures are exerted on the surface of a silicon wafer through contact-loading. It is well recognized that metallic β-tin (Si-II) phase is generally formed at a hydrostatic pressure between 10 and 13 GPa, which may be followed with the transformation to other metastable Si-XII, Si-III or/and amorphous Si (α-Si) phases during the subsequent decompression process depending on the unloading rate. It is generally believed that the imposed pressure and its release rate predominantly dictate the phase transformation of Si. However, how the inherent properties such as conduction type and carrier concentration affect the pressure-induced phase transformation of Si remains unclear. Here, we experimentally unveil that the increasing hole concentration facilitates the amorphisation in the micro-indented Si with a constant unloading rate. Moreover, based on the establishment of reliable interatomic potentials for the doped Si using machine learning, we theoretically reveal that during the decompression process starting from Si-II phase the holes are preferentially localized around the three-coordinated (CN = 3) Si atoms, which are therefore stabilized. In this context, the higher concentration of holes enable the formation of more stabilized CN = 3 Si atoms, which facilitates the amorphisation of Si. Such a theoretical work sheds light on the intricate relationship between electronic structure and structural dynamics of Si, offering valuable insights into the pressure-induced phase transformation of Si. Furthermore, the presence of high enough concentration of holes is also experimentally confirmed to facilitate the indentation-induced amorphisation in other prominent semiconductors such as Ge, GaAs and SiC. Our work discovers that the hole concentration is even another determining factor for the pressure-induced phase transformations of the industrially important semiconductors. Such a discovery bears technological implication for the precision wafering of the aforementioned semiconductors, in which the pressure-induced phase transformation may readily occur.

Pressure-induced phase transformation of silicon (Si) ubiquitously occurs in most wafering processes such as slicing, grinding and lapping, where high pressures are exerted on the surface of a Si wafer through contact-loading[1-5]. As the most important and widely used semiconductor, Si has been intensively investigated in the regard of pressure-induced phase transformation for the past decades by means of indentation, which is the most convenient and standard way for the scientific evaluation of all contact-loading related phenomena[4-6]. To date, it has been well known that Si transforms from the diamond cubic (Si-I) structure to the metallic β-tin (Si-II) structure at hydrostatic pressures of 10-13 GPa[6-11]. On release of pressure, Si-II phase transforms to other metastable phases Si-XII (r8), Si-III (bc8) or/and amorphous Si (α-Si) depending on the unloading rate[12-19]. The previous studies usually focused on the effects of indentation conditions such as the unloading rate, indenter angle and the maximum load on the phase transformation of Si[12, 14, 20-23]. To date, it has not been addressed how the inherent properties such as charge carrier type and concentration affect the pressure-induced phase transformation of Si, which is not only of scientific curiosity but also of practical significance because Si should be usually doped to be n- or p-type with specific carrier concentrations for the technological applications. To the best of our knowledge, only Yan *et al.* reported that the formation of α-Si phase in nano-indented Si was significantly promoted by the heavy boron (B)-doping with a concentration of 9 × $10^{18}$ cm$^{-3}$.[24] However, the underlying mechanism was not essentially understood. Tentatively, they suggested that the elastic strain arising from the large size mismatch between B and Si atoms, together with the reduction of shear elastic constant caused by the heavy B-doping, played important roles in the facilitated amorphisation[24]. Evidently, the exact mechanism underlying the effect of heavy B-doping on the pressure-induced phase transformation of Si needs to be further explored.

In this work, we investigated the phase transformation behaviors in various doped Si slices, which were subjected to micro-indentations by a diamond Vickers micro-indenter. We found that the formation of α-Si phase in the micro-indented Si is continuously facilitated by increasing the doped B concentration ([B]) from ~5 × $10^{17}$ to 1 × $10^{19}$ cm$^{-3}$. More importantly, we unveiled that the aforementioned facilitated amorphisation due to the increase in [B] is essentially attributed to the increase in hole concentration, not the B-doping modified mechanical properties as supposed previously[24]. The mechanism underlying the unveiled hole-facilitated amorphisation in the pressure-induced phase transformation of Si is explored by the theoretical calculations based on the machine-learning interatomic potentials of the doped Si. As an extension of knowledge, our work demonstrates that the hole-facilitated amorphisation also occurs in the pressure-induced phase transformation of Ge, GaAs or SiC.

The pressure-induced phase transformation behaviors in lightly ($8.04 \times 10^{14}$ cm$^{-3}$) and heavily ($1.19 \times 10^{19}$ cm$^{-3}$) B-doped Si slices are manifested by the Raman spectra (Fig. 1a and b). In every spectrum, the peak at 521 cm$^{-1}$ originates from the Si-I matrix surrounding the deformed contact region within the micro-indentation[12, 14, 21, 22]. The peaks at 382 and 435 cm$^{-1}$ are assigned to Si-III phase and those at 350, 394 and 485 cm$^{-1}$ to Si-XII phase[12, 25-27]. While, the broad band centered at 475 cm$^{-1}$ arises from α-Si phase[28-31]. In Fig. 1a or 1b, the 30 Raman spectra are categorized into three groups: sole α-Si (in red), α-Si + Si-III + Si-XII (in green) and Si-III + Si-XII (in blue). Such presence and distribution of residual phases are due to the crystallographic anisotropy and inhomogeneity of the stress distribution within the micro-indentations[18, 32]. For the sake of quantitative analysis, we define the percentage of the micro-indentations featuring sole α-Si phase in the 30 micro-indentations as the amorphisation probability, denoted as $P_{\alpha\text{-Si}}$, hereafter. Namely, $P_{\alpha\text{-Si}} = \beta/30 \times 100\%$, where $\beta$ is the number of the micro-indentations with sole α-Si phase. Hence, the values of $P_{\alpha\text{-Si}}$ for the lightly and heavily B-doped Si slices are calculated to be 20% and 60%, respectively. Clearly, the heavily B-doped Si is more readily transformed into α-Si phase than the lightly B-doped Si when subjected to the micro-indentations.

The heavy B-doping leads to an appreciable lattice tensile strain because the covalent radius of B atom (0.88 Å) is much smaller than that of Si atom (1.17 Å)[24, 33]. In order to examine the effect of such strain on the pressure-induced phase transformation of Si, we co-doped germanium (Ge) with a covalent radius of 1.22 Å into heavily B-doped Si (Si:B) to offset the lattice tensile strain[34]. Figure 1c shows the similar values of $P_{\alpha\text{-Si}}$ around 60% for the Si:B slice and the three heavily B and Ge-codoped Si [Si:(B,Ge)] slices with different concentrations of B and Ge. Herein, $P_{\alpha\text{-Si}}$ is also derived from the 30 micro-indentations imposed on each slice. According to Vegard's law[35], as the ratio of Ge concentration ([Ge]) to [B] ([Ge]/[B]) reaches ~7.2, the lattice tensile strain induced by the B-doping can be completely offset by the Ge-codoping. In Fig. 1c, the Si:(B,Ge)-1, 2 and 3 slices respectively have the [Ge]/[B] ratios of 4.31, 5.20 and 5.25, indicating the increasingly smaller lattice tensile strain in order. Meanwhile, such three Si:(B,Ge) slices have nearly the same $P_{\alpha\text{-Si}}$ as that of Si:B counterpart (Fig. 1c). Therefore, the facilitated amorphisation in the micro-indented Si due to the heavy B-doping is definitely not relevant to the lattice tensile strain.

Figure 1d shows the values of $P_{\alpha\text{-Si}}$ for the Si:B slice, heavily B and phosphorus (P)-codoped Si [Si:(B,P)] and heavily P-doped Si (Si:P) slices, each of which was imposed with 30 micro-indentations. The Si:(B,P)-1 slice is codoped with a [B] of $1.28 \times 10^{19}$ cm$^{-3}$ and a P concentration ([P]) of $1.23 \times 10^{19}$ cm$^{-3}$, resulting in a net hole concentration of ~$5.0 \times 10^{17}$ cm$^{-3}$, which is much smaller than that (~$1.2 \times 10^{19}$ cm$^{-3}$) of the Si:B slice. Noteworthily, the $P_{\alpha\text{-Si}}$ of Si:(B,P)-1 slice is only ~33%, much smaller than that (60%) of the Si:B slice. For the Si:(B,P)-2 slice codoped with a [B] of $1.35 \times 10^{19}$ cm$^{-3}$ and a [P] of $3.69 \times 10^{19}$ cm$^{-3}$, which results in n-type conduction with a net electron concentration of ~$2.34 \times 10^{19}$ cm$^{-3}$, the $P_{\alpha\text{-Si}}$ is also only ~30%. The above results indicate that $P_{\alpha\text{-Si}}$ decreases significantly as the high hole concentration induced by the heavy B-doping is significantly offset by the P-codoping. For reference, the Si:P slice with an electron concentration of ~$6.52 \times 10^{19}$ cm$^{-3}$ has a smaller $P_{\alpha\text{-Si}}$ value of ~23%. Consequently, we conclude that the presence of a high concentration of holes is the root cause for the facilitated amorphisation in the micro-indented Si due to the heavy B-doping. In fact, this is also the case for the micro-indented Si with heavy gallium (Ga)-doping, which also leads to a high hole concentration (Extended Data Fig. 1). Meaningfully, our extended investigations on the pressure-induced phase transformation of Ge, GaAs or SiC also unveil that the presence of a sufficiently high concentration of holes leads to the facilitated amorphisation (Extended Data Fig. 2 to 4). It is a common sense that the hole concentration dictates the electrical conductivity (resistivity) of a p-type semiconductor. While, in this work we discover that the hole concentration is even closely correlated with the structural evolution in the phase transformation behaviors of a variety of p-type semiconductors, raising an intriguing scientific issue.

In this work, we tentatively focus on the origin of the hole-facilitated amorphisation in the pressure-induced transformation of Si. To this end, we performed molecular dynamics (MD) simulations based on the reliable interatomic potentials of the doped Si achieved by machine learning to systematically investigate the order-disorder phase transitions in the decompressed Si-II systems with different hole concentrations. The well-trained potentials can predict the atomic forces pertaining to the static and dynamic properties of Si with density functional theory (DFT)-level accuracy (Extended Data Fig. 5). More importantly, such well-trained potentials are competent to well simulate the pressure-induced phase transformation of Si (Extended Data Fig. 6).

The structural evolution during the decompression process starting from Si-II phase is analyzed in terms of the bond-orientational order (BOO) parameter, where the local structure ordering is defined with a value of $Q_6$[36]. The Si atoms with $Q_6 \leq 0.2$ are classified as amorphous ones and $Q_6 \geq 0.3$ as crystalline ones. As shown in Fig. 2a, under ambient pressure the Si atoms with $Q_6 \leq 0.2$ (α-Si atoms) increase with the hole concentration. Figure 2b shows the fraction of α-Si atoms as a function of the hole concentration when the system is decompressed to ambient pressure. Clearly, the fraction of α-Si atoms increases remarkably when the hole concentration exceeds a threshold value. This trend is highly reflected by the experimental result that illustrates the dependence of amorphisation probability on the hole concentration for the micro-indented Si (Fig. 2c). Note that the decompression rate is several orders of magnitude higher in the calculations than in the experiments, and, a much higher decompression rate enables Si-II phase to much more readily transform into α-Si[12, 14]. Thus, it is understandable that the calculated fraction of α-Si atoms (Fig. 2b) is much larger than the experimentally derived amorphisation probability (Fig. 2c) at a given hole concentration. Moreover, along with the increasing hole concentration the fraction of α-Si atoms increases synchronously with that of three-coordinated (CN = 3) Si atoms (Fig. 2b and Extended Data Fig. 7), in accord with the common sense that the existence of CN = 3 Si atoms is detrimental for the crystallinity of Si. In view of the above results, we know that during the decompression process starting from Si-II phase, more CN = 3 Si atoms are formed due to the presence of a higher concentration of holes, leading to the facilitated amorphisation of Si.

In order to clarify the effect of the hole concentration on the local electronic states and the structural ordering during the decompression process starting from Si-II phase, we studied the density of states (DOS) projected on each Si atom in the two systems with the hole concentrations of 0 and 0.1 at.%, respectively. Since the carriers mainly affect the electronic states around the Fermi level ($E_F$)[37], the evolutions of DOS at $E_F$ (DOS@$E_F$) during the decompression process are presented for the CN = 3, 4 and 5 Si atoms, respectively (Fig. 3a and b). Herein, all the CN = 3, 4 and 5 Si atoms account for ~98% of the total Si atoms in a calculated system. During the decompression process, the number of CN = 3 Si atoms increases while that of CN = 5 Si atoms decreases gradually (Fig. 3a and b). More importantly, the distribution curve of DOS@$E_F$ for the CN = 3 Si atoms in the system with the hole concentration of 0.1 at.% is largely broadened and shifts to the larger DOS values (Fig. 3b), with respect to that for the CN = 3 Si atoms in the system without holes (Fig. 3a). While, the distribution curves of DOS@$E_F$ for the CN = 4 or 5 Si atoms in the two systems shift little in terms of the DOS values (Fig. 3a and b). Therefore, the presence of holes mainly affects the electronic states of the CN

= 3 Si atoms. Since the valence state of the CN = 3 Si atoms is higher in energy than those of the CN = 4 and 5 Si atoms (Extended Data Fig. 8), it can be derived that the holes are preferentially localized around the CN = 3 Si atoms in the system with the hole concentration of 0.1 at.%. Such localization of holes can also be manifested by the correlation between the electron-deficient Si atoms and the CN = 3 ones. As shown in Fig. 3c and d, the atoms with Mulliken charge < 3.9e (Mulliken charge = 4e for perfect Si crystal) are mainly the CN = 4 Si atoms in the absence of holes, while those are mainly the CN = 3 Si atoms in the presence of holes with the concentration of 0.1 at.%. Accordingly, we know that the localization of holes favors the stabilization of the CN = 3 Si atoms, and, the presence of a high concentration of holes results in more CN = 3 Si atoms formed during the decompression process starting from Si-II phase.

To further understand the effect of hole concentration on the amorphisation in the indented Si, we performed isobaric simulations for the initial 100 ps of the phase transformation starting from Si-II phase for the aforementioned two systems. As shown in Fig. 4a, in the absence of holes the coordination of the tracked Si atom fluctuates between CN = 3 (manifested by the maximum bond length > 3.0 Å) and CN = 4 (manifested by all the four bond lengths < 3.0 Å), accompanying with the valence charge varying around 4e, which indicates the well stability of CN = 4 coordination. While, in the presence of holes (0.1 at.%) the coordination of the tracked Si atom changes from CN = 4 to CN = 3 irreversibly, meanwhile, the valence charge becomes successively smaller than 4e, indicating the hole localization around the CN = 3 Si atom. Addressing the four covalent bonds of the tracked Si atom, we define the difference between the maximum and minimum bond-lengths as Δbond-length. As shown in Fig. 4b, the Δbond-length oscillates constantly with the elapse of time in the absence of holes. While, in the presence of holes (0.1 at.%), the Δbond-length increases instantly and tends to be stable around quite a large value, indicating that the CN = 3 Si atom is dynamically stabilized due to the hole localization as mentioned above. Furthermore, for each Si atom, we define an order parameter ($\xi$) as the ratio of Δbond-length to ⟨bond-length⟩ that represents the average bond length of the four covalent bonds. Fig. 4c shows the evolutions of distribution curves of $\xi$ for all the atoms with Mulliken charge < 3.9e in the two systems during the isobaric simulations. In the absence of holes, the $\xi$-distribution curves change slightly during the isobaric process, suggesting the well-remained CN = 4 coordination. Whereas, in the presence of holes (0.1 at.%), the main peak of the $\xi$-distribution curve shifts from $\xi \approx 0.05$ to 0.45 successively along with the isobaric process, indicating the irreversible formation of more CN = 3 Si atoms. Again, it can be derived from Fig. 4 that the increase in hole concentration facilitates the formation of dynamically stabilized CN = 3 Si atoms during the decompression process starting from Si-II phase, leading to the irreversible destruction of the long-range order of crystalline Si and therefore the amorphisation of Si.

Our experimental study has unveiled the hole-facilitated amorphisation in the pressure-induced transformation of Si. On the other hand, our theoretical study has successfully developed the reliable interatomic potentials for the doped Si through machine learning. On this basis, it is found that during the decompression process starting from Si-II phase the CN = 3 Si atoms are dynamically stabilized by the doped holes. In this context, the higher concentration of holes lead to the formation of more dynamically stabilized CN = 3 Si atoms thus facilitating the amorphisation of Si. Such a theoretical work sheds light on the intricate relationship between the electronic structure and the structural dynamics of Si, offering valuable insights into the pressure-induced phase transformation of Si. We have also found that the presence of a sufficiently high concentration of holes facilitates amorphisation in the pressure-induced phase transformation of Ge, GaAs or SiC. This work discovers that the hole concentration is even another determining factor for the pressure-induced phase transformations of the industrially important semiconductors, bearing not only the scientific significance in semiconductor physics but also the technological implication for the precision wafering of such semiconductors.

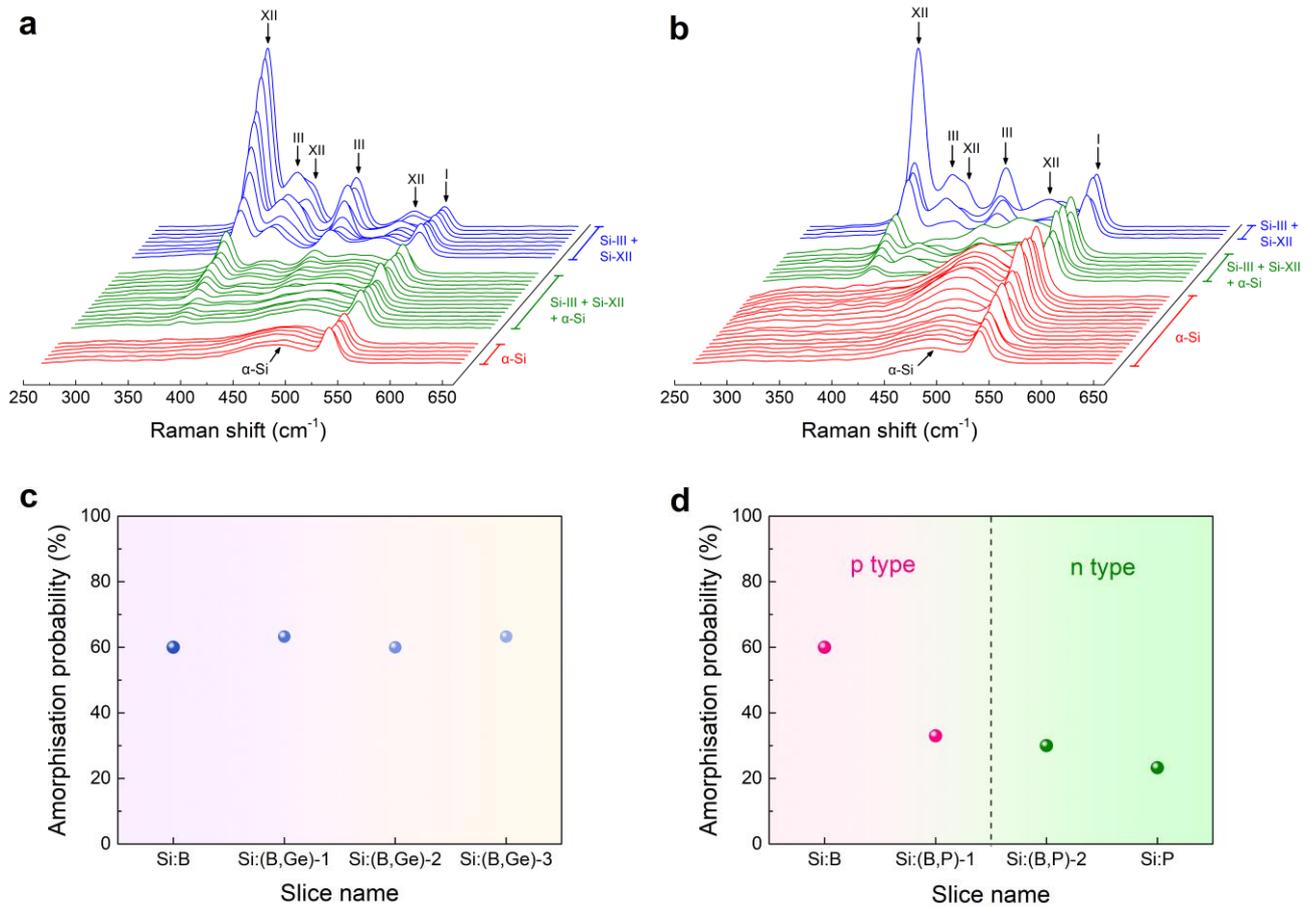

**Fig.1| The pressure-induced phase transformation behaviors in a variety of doped Si slices. a,** Raman spectra acquired from the 30 micro-indentations imposed on lightly B-doped Si slices ([B] = 8.04 × $10^{14}$ cm$^{-3}$). **b,** Raman spectra acquired from the 30 micro-indentations imposed on heavily B-doped Si slices ([B] = 1.19 × $10^{19}$ cm$^{-3}$). **c,** The amorphisation probability for each of the micro-indented Si slices with different B and Ge concentrations, including: Si:B slice (solely B-doped, [B] = 1.19 × $10^{19}$ cm$^{-3}$), Si: (B,Ge)-1 slice (B and Ge-codoped, [B] = 1.28 × $10^{19}$ cm$^{-3}$ and [Ge] = 5.52 × $10^{19}$ cm$^{-3}$), Si: (B,Ge)-2 slice (B and Ge-codoped, [B] = 1.45 × $10^{19}$ cm$^{-3}$ and [Ge] = 7.54 × $10^{19}$ cm$^{-3}$) and Si:(B,Ge)-3 slice (B and Ge-codoped, [B] = 1.57 × $10^{19}$ cm$^{-3}$ and [Ge] = 8.25 × $10^{19}$ cm$^{-3}$). **d,** The amorphisation probability for each of the micro-indented Si slices with different B and P concentrations, including: Si:B slice (solely B-doped, [B] = 1.19 × $10^{19}$ cm$^{-3}$), Si:(B,P)-1 slice (B and P-codoped, [B] = 1.28 × $10^{19}$ cm$^{-3}$ and [P] = 1.23 × $10^{19}$ cm$^{-3}$), Si:(B,P)-2 slice (B and P-codoped, [B] = 1.35 × $10^{19}$ cm$^{-3}$ and [P] = 3.69 × $10^{19}$ cm$^{-3}$) and Si:P slice (solely P-doped, [P] = 6.52 × $10^{19}$ cm$^{-3}$).

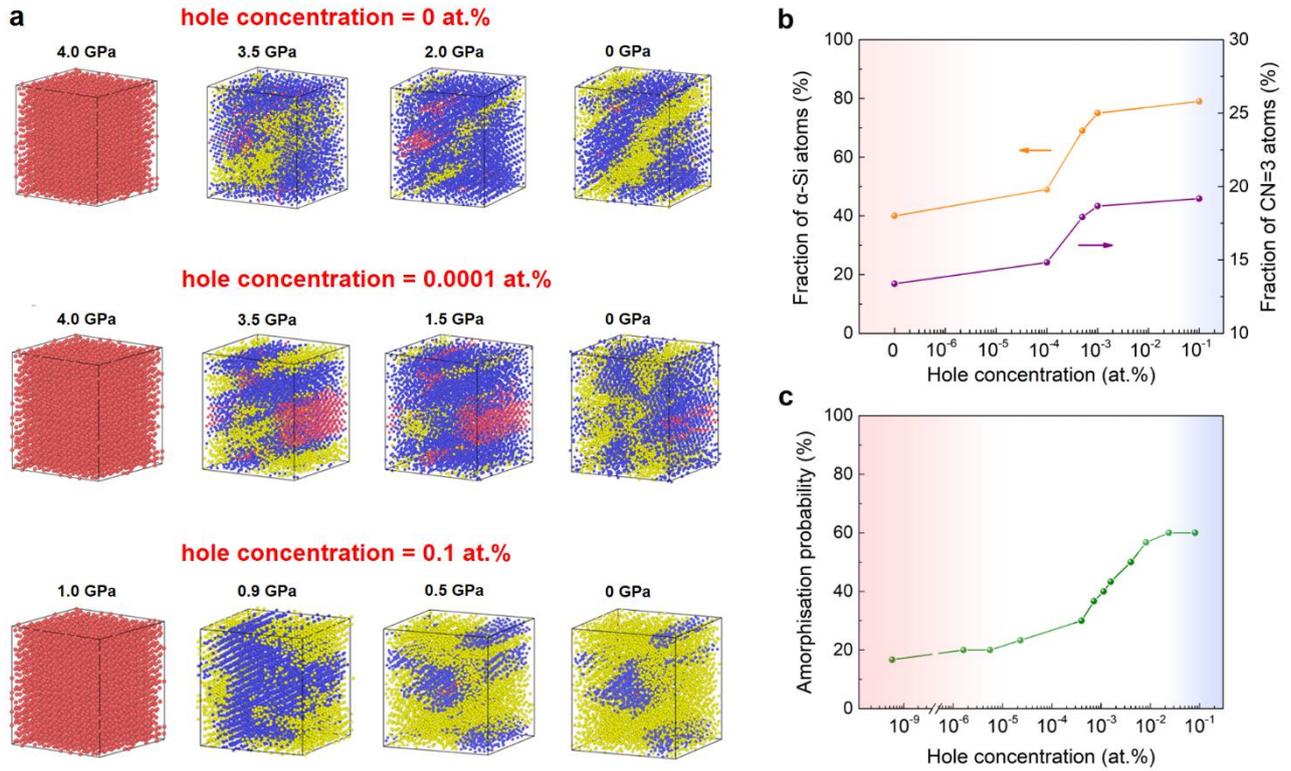

**Fig. 2| Structural evolution during the decompression process starting from Si-II phase. a,** Snapshots of the atomic configurations during the decompression process for the three systems with different hole concentrations. The spatial distribution of the atoms with high $Q_6 \geq 0.3$ (in red), medium $Q_6$ ($0.2 \leq Q_6 \leq 0.3$) (in blue) and low $Q_6 \leq 0.2$ (in yellow) values are labelled, respectively. All structural drawings were created using OVITO[37-39]. **b,** The fractions of α-Si atoms (in orange) and three-coordinated (CN = 3) Si atoms (in purple) in the systems with different hole concentrations under ambient pressure. **c,** Experimental result illustrating the dependence of the amorphisation probability on the hole concentration for the micro-indented Si.

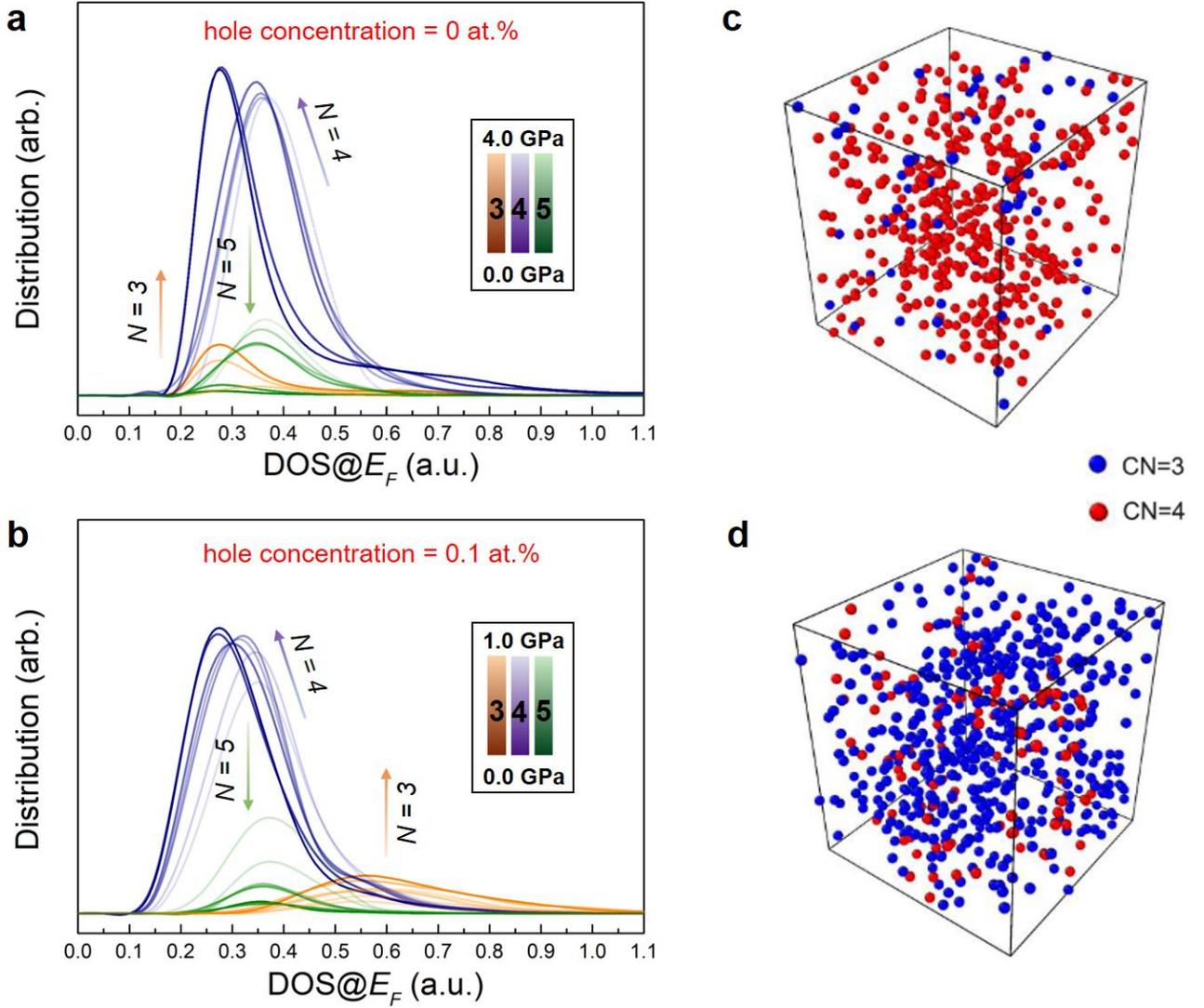

**Fig. 3| Evolutions of the density of states at Fermi level (DOS@$E_F$) for the Si atoms with different coordination numbers during the decompression process starting from Si-II phase and the hole distribution under ambient pressure in the two systems with the hole concentrations of 0 and 0.1 at.%, respectively. a, b,** Evolutions of DOS@$E_F$ for the CN = 3, 4 and 5 Si atoms during the decompression process. The orange, green and purple curves correspond to the CN = 3, 4 and 5 Si atoms, respectively, and the color of each kind of curves gradually becomes darker along with the decompression process. The arrows indicate the direction of evolution of the curves with decreasing pressure. **c, d,** The Si atoms with Mulliken charge < 3.9e are shown for the two systems under ambient pressure, where the CN = 3 and 4 Si atoms are colored blue and red, respectively. The minority CN = 5 Si atoms are not shown.

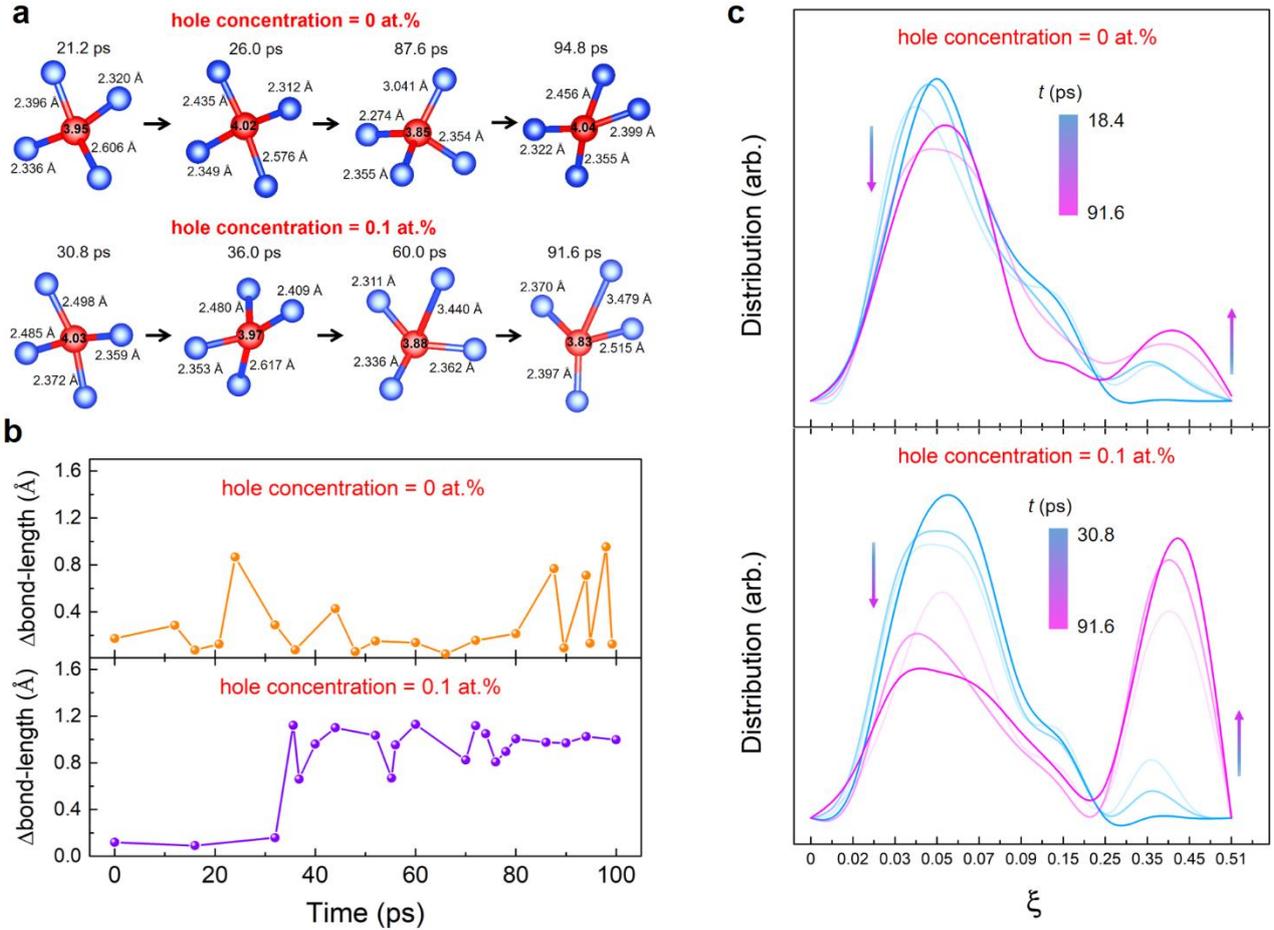

**Fig. 4| Evolutions of the local structure during the initial 100 ps of the phase transformation starting from Si-II phase in the two systems with the hole concentrations of 0 and 0.1 at.%, respectively. a,** The changes of four covalent bond lengths along with the isobaric time for the tracked Si atom (in red). The number labeled on the tracked atom is the amount of valence charges (in a unit of e). **b,** Δbond-length (the difference between the maximum and the minimum bond lengths for the four covalent bonds of the tracked Si atom) as a function of isobaric time. **c,** Evolutions of order parameter (ξ) distribution for the two systems during the isobaric process. The arrows indicate the direction of evolution of the curves with increasing isobaric time.

## Methods

### Micro-indentation

To investigate the pressure-induced phase transformation behaviors of the semiconductors as mentioned above, 30 micro-indentations without lateral cracks were imposed on the polished surface of each Si, Ge, GaAs or SiC slice by using a diamond Vickers micro-indenter (Struers DuraScan 10). Here, all the imposed slices were sized in $2.0 \times 2.0$ cm$^2$. The specifications of the aforementioned semiconductor slices are given in Extended Data Tab.1-7. The Si, Ge, GaAs and SiC slices were imposed with the peak loads of 500, 250, 250 and 2000 mN, respectively, with a loading time of 5 s. After holding at the peak load for 30 s to minimize the time-dependent plastic effect, the indenter was unloaded immediately.

### Micro-Raman spectroscopy characterization

Micro-Raman spectra were immediately acquired for each slice imposed with the aforementioned 30 micro-indentations by using a micro-Raman spectrometer (Bruker Senterra), where a 50 ×, long working distance objective lens with a numerical aperture of 0.75 was used. To acquire a Raman spectrum, a 532 nm laser beam was focused onto the central region of a micro-indentation as determined by optical microscopy. Moreover, the size of laser spot was about 2 μm, which is smaller than the dimensions of a micro-indentation, and the laser power was kept at a level of 20 mW.

### Molecular-dynamics simulations

Molecular-dynamics (MD) simulations for the 5832-atom systems were carried out using LAMMPS[40]. A Nosé-Hoover thermostat was used to control the temperature and a Parrinello-Rahman barostat, integrated with the method of Martyna et al. was used to control the hydrostatic pressure after the energy minimization[41-43]. The initial velocities of the atoms were assigned according to the Maxwell-Boltzmann distribution. The deep potential force field for different hole concentrations (0, 0.0001, 0.0005, 0.001 and 0.1 at.%) was generated using DeePMD-kit[44, 45]. For different hole concentrations in the crystalline Si-II structure, decompression simulations from 12 GPa to ambient pressure were performed at 300 K with a decompression rate of 0.02 GPa/ps. Isothermal-isobaric ensemble (NPT) simulations were conducted for 10 ps after every decompression of 2 GPa. An additional NPT simulation was carried out for the initial 100 ps of phase transformation starting from Si-II phase for the two systems with the hole concentrations of 0 and 0.1 at.%, respectively. A Velocity Verlet integrator was employed for the decompression and the isobaric simulations with the time steps of 2 and 1 fs, respectively.

**Deep Potential training**

**Reference dataset**. To build a general neural network potential capable of describing the crystalline and amorphous phases during the decompression process, we constructed a DeePMD model training database. This database includes Si-I, Si-II, Si-III, Si-IV, α-Si phases with different hole-concentrations (0, 0.0001, 0.0005, 0.001 and 0.1 at.%). The α-Si phase configuration was obtained by heating Si-I from room temperature (RT) to its melting point, followed by an isothermal and isobaric simulation at RT under a given pressure ranging from 0 to 12 GPa for 300 fs. To ensure that our DeePMD model has robust extrapolation capabilities for the large-scale configurations, we used different system sizes as follows.

Si-I/Si-III: 16, 32, 64, 96 and 128 atoms

Si-II: 16, 32, 64, 96, 108, 128 and 216 atoms

Si-IV: 24, 48, 72, 96, 144 atoms

α-Si: 96, 128 atoms.

The total database comprised about 68000 different atomic configurations. The optimized crystalline structures and potential energies for all the configurations in the database were mainly populated from ab initio molecular dynamics (AIMD) trajectories employing the Perdew-Burke-Ernzerhof (PBE) exchange-correlational functional[46]. The electron-nucleus interactions were described using the Ultrasoft pseudopotentials with 4 valence electrons for Si and the plane wave basis cut-off energy was set to 250 eV [47]. The canonical ensemble (NVT) was maintained at 300 K with the Nosé-Hoover thermostat[48]. All AIMD simulations were performed with a timestep of 1 fs, employing a relation time of 300 fs.

**Active learning of deep potential.** The deep potential generator (DP-GEN) was utilized to explore new data and to train the DeePMD model. A complete iteration includes DeePMD training, MD exploration, and DFT re-evaluation. The DeePMD-kit program developed by Zhang *et al* was used to train the neural network potential adopting the DeepPot-SE (Deep Potential-Smooth Edition) approach[49, 50]. The details of this method can be found in ref. 51. The model includes embedding and fitting networks. The embedding network includes three hidden layers with 25, 50 and 100 neurons, respectively. The fitting network contains three hidden layers with 240 neurons per layer. A smooth-edition, angular descriptor was used to extract the local environment of particles. We used a cutoff radius of 6 Å for searching neighbor with smoothing starting from 5.8 Å. The deep neural network was optimized using Adam stochastic gradient descent method, with an initial learning rate of 0.001 that decreases exponentially every 5000 steps. The loss function is defined by

$$L = \frac{p_e}{N}\Delta E^2 + \frac{p_f}{3N}\sum_i |\Delta \boldsymbol{F}_i|^2$$

where $\Delta E$ and $\Delta \boldsymbol{F}_i$ are the differences between the training data and current deep potential prediction for the atomic energy and force, respectively. $N$ is the number of atoms. The prefactor $p_e$ increase from 0.02 to 1 eV$^{-2}$ and $p_f$ decayed from 1000 to 1 Å$^2$eV$^{-2}$. Four functionally equivalent DeePMD models with different initial random seeds for the stochastic neural network training algorithm were trained based on the aforementioned dataset. These four DeePMD models were used to perform several short classical MD with 300 steps under the NPT conditions. The NPT simulations were conducted at temperatures of 200, 300, and 400 K and pressures of 1, 3, 5, 7, 9, and 11 GPa to fully explore the configurational space of the five Si phases. During the simulation, the atomic forces were predicted by the aforementioned four neural network DeePMD models simultaneously. We assessed model convergence using the so-called maximum standard force deviation $\delta F$, which was calculated by the following expression:

$$\delta F = \max_i \left( \sqrt{\left\langle \left| \boldsymbol{F}_{w,i}(\mathcal{R}_t) - \langle \boldsymbol{F}_{w,i}(\mathcal{R}_t) \rangle \right|^2 \right\rangle} \right)$$

where $\boldsymbol{F}_{w,i}$ is the force acting on $i^{th}$ atom predicted by the deep potential model with the parameter of $w$, and $\mathcal{R}_t$ is the atomic position. If the values of $\delta F$ predicted by the aforementioned four models are in a specific range of 0.15-0.35 eV/Å, the corresponding configuration will be re-evaluated and labeled by the single-point DFT calculations using VASP5.4.4[52] and then supplemented into the dataset for the next training. During the re-evaluated DFT calculations, the PBE exchange-correlation functional and Ultrasoft pseudopotentials were used with an energy cutoff of 250 eV. The updated dataset was continually trained using the DeePMD potential with the parameters same as those used before so as to enable the predictions from the aforementioned four models to be always consistent.

**Accuracy of the DeePMD model.** The predictive power of the DeePMD model is shown in Extended Data Tab.8 and Extended Data Fig. 5. Extended Data Tab.8 contains information on the root-mean-square errors (RMSEs) in energies and forces for training and testing datasets, serving as the indicators to evaluate the model. For all DeePMD models, RMSEs for the training and testing sets are close to each other in terms of energies and forces. Moreover, the RMSEs in the atomic energies are below 6.12 and 5.18 meV/atom for the training and testing sets, respectively. This demonstrates that the six DeePMD potentials of Si with different hole concentrations can accurately predict the potential energies. The atomic forces predicted by the DeePMD model are highly consistent with the results of DFT calculations, as confirmed by the correlation coefficients, $R^2$, for the crystalline and amorphous Si phases (Extended Data Fig. 5). Therein, the $R^2$ values exceed 0.997 and 0.993 for Si-II and α-Si phases, respectively, even better than the Gaussian approximation potential (GAP) results. In contrast, the Tersoff potentials fail to predict the behaviors of Si-II and α-Si phases accurately, as manifested by the considerably low correlation coefficients. Thus, we conclude that our DeePMD models can well describe the ab initio potential energy surface and atomic force for Si-II and α-Si phases.

**Self-consistent calculations and projected DOS**

The real-space Kohn-Sham density functional theory (KS-DFT) method, RESCU was used to calculate the electronic structures of such 5832-atom systems. The technical details of RESCU can be found in the original literature[53]. In the KS-DFT self-consistent calculation, a Γ- centered $1 \times 1 \times 1$ Monkhorst-Pack grid was used for the configuration obtained in the MD simulation. The electron exchange-correlation was treated by Perdew-Burke-Ernzerhof (PBE) exchange-correlational functional and a linear combination atomic orbital (LCAO) basis set was used to capture the Kohn-Sham states. The self-consistent electronic loop will not stop until the total energy reaches the tolerance of $10^{-5}$ eV.

**Data availability**

All data are available in the manuscript or the supplementary materials.

**Acknowledgements:** We thank Prof. X.Chen with State Key Laboratory of Crystal Materials, Shandong University, China for her providing semiconductor SiC wafers. This work was supported by the National Natural Science Foundation of China (62174145 and 62090031), National Key R&D Program of China (2019YFE0112000) and Natural Science Foundation of Zhejiang Province (LR21A040001).

**Author contributions:** X.Ma., Y.Lu., and D.Yang. conceived and supervised the project. X.Ma. designed all experiments. Y.Lu. designed all theoretical calculations. T.Zhao., and Y.Sun. performed the micro-indentations and Raman spectroscopy measurements. D.Wu., and H.Chen.  performed resistivity and carrier concentration measurements. T.Zhao and Y.Sun. performed experimental data analyses. S.Zhong. performed DFT calculations and molecular-dynamics simulations. C. Zhang., R.Shi., Z.Ni., and X.Pi. provided discussions. X.Ma., Y.Lu., T.Zhao. and S.Zhong. wrote the paper with input from all other authors.

**Competing interests:** The authors declare no competing interests.

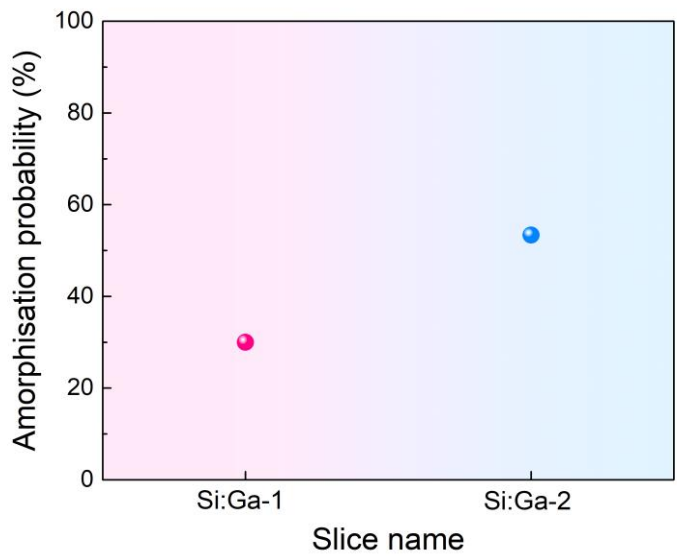

**Extended Data Fig. 1| The amorphisation probability for lightly or heavily Ga-doped Si slice subjected to 30 micro-indentations.** The lightly and heavily Ga-doped Si slices, denoted as Si:Ga-1 and Si:Ga-2, contain Ga concentrations of 3.60 × $10^{16}$ and 1.67 × $10^{18}$ cm$^{-3}$, respectively. The amorphisation probability for the heavily Ga-doped slice is significantly larger than that for the lightly Ga-doped slice, further supporting that the facilitated amorphisation in the indented Si is ascribed to the high hole concentration.

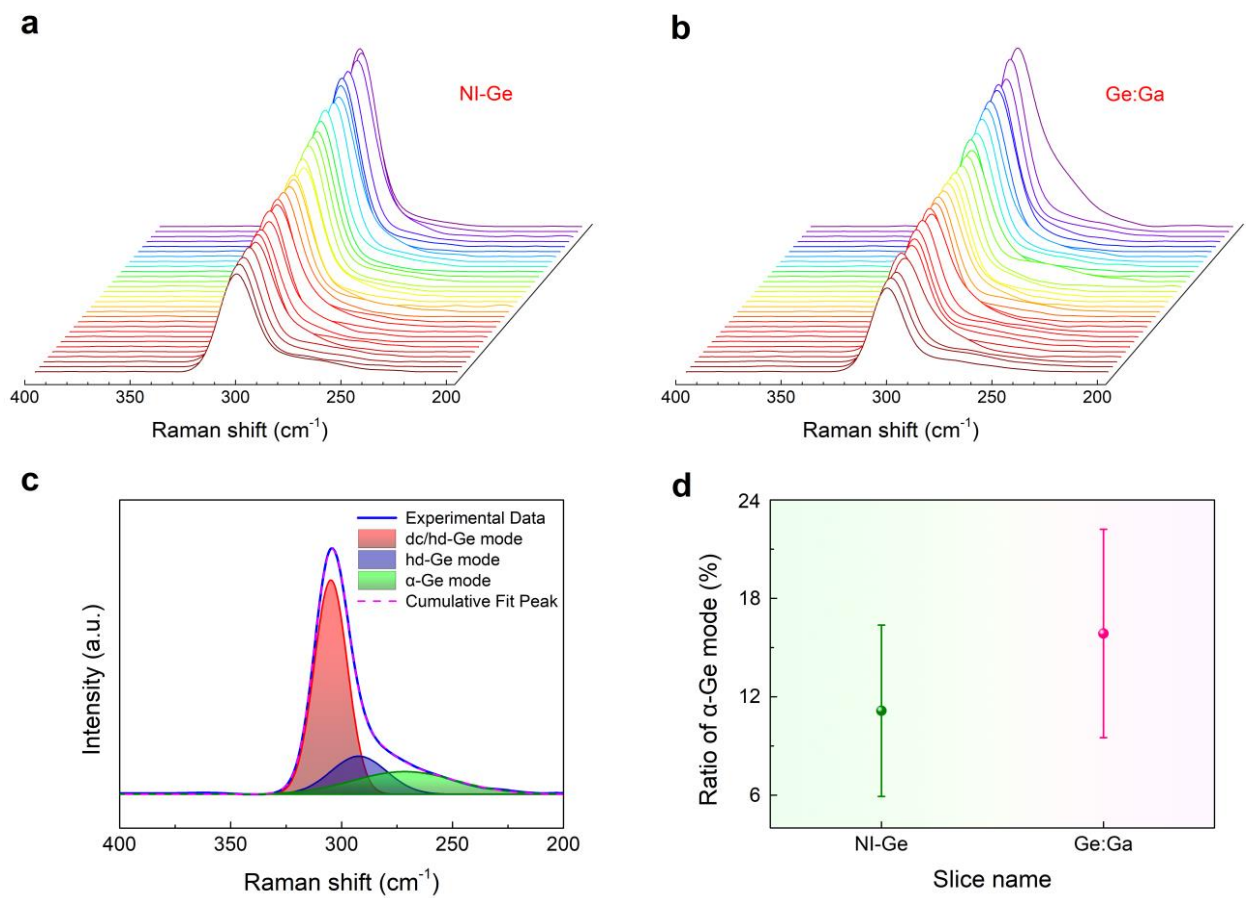

**Extended Data Fig. 2| The pressure-induced phase transformation behaviors in various Ge slices. a,** Raman spectra acquired from the 30 micro-indentations imposed on the near-intrinsic Ge (NI-Ge) slice. **b,** Raman spectra acquired from the 30 micro-indentations imposed on the heavily Ga-doped Ge (Ge:Ga) slice with a hole concentration of $2.08 \times 10^{18}$ cm$^{-3}$. **c,** Deconvolution of a representative Raman spectrum of the indented Ge slice, manifesting with a broad amorphous Ge (α-Ge) band peaking at ~270 cm$^{-1}$ (in green), a hd-Ge band peaking at ~290 cm$^{-1}$ (in blue), and a pronounced band peaking at ~300 cm$^{-1}$ corresponding to the mixture of hd-Ge and dc-Ge phases (in red)[54-56]. **d,** Statistical results of the ratios of α-Ge mode in the Raman spectra shown in **a** and **b**. Herein, the ratio of α-Ge mode in a Raman spectrum is defined as the ratio of the integrated intensity of α-Ge band to that of the full spectrum. Note that the average ratio of α-Ge mode for the Ge:Ga slice is much larger than that for the NI-Ge slice, indicating that the amorphisation in the indented Ge is facilitated by the presence of a higher concentration of holes.

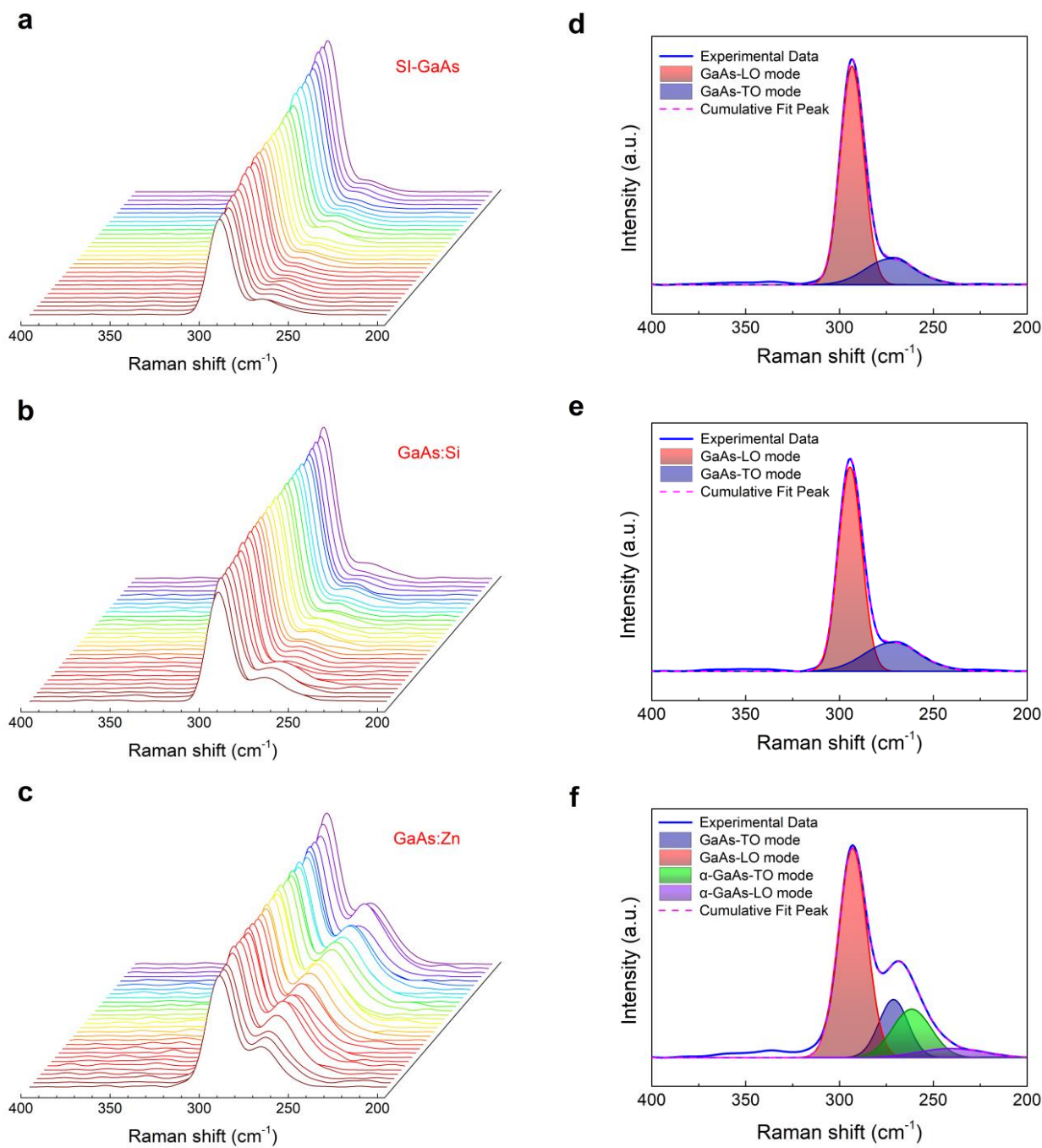

**Extended Data Fig. 3| The pressure-induced phase transformation behaviors in various GaAs slices. a-c,** Raman spectra acquired from the 30 micro-indentations imposed on the semi-insulating GaAs (SI-GaAs), heavily Si-doped GaAs (GaAs:Si, n-type) slice with an electron concentration of 1.31 × $10^{18}$ cm$^{-3}$, and heavily Zn-doped GaAs (GaAs:Zn, p-type) slice with a hole concentration of 1.48 × $10^{19}$ cm$^{-3}$, respectively. **d-f,** Representative Raman spectra and their deconvolutions for the indented SI-GaAs, GaAs:Si, and GaAs:Zn slices, respectively. Both the Raman spectra of the indented SI-GaAs and n-type GaAs:Si slices can be deconvoluted into two sub-bands for GaAs-LO mode peaking at ~294 cm$^{-1}$ (in red) and GaAs-TO mode peaking at ~270 cm$^{-1}$ (in blue), which is in accordance with the experimental results reported in Refs. 57 and 58. While, for the deconvoluted Raman spectrum

of the indented p-type GaAs:Zn slice, besides the two sub-bands of GaAs-LO and GaAs-TO modes, another two sub-bands of amorphous GaAs (α-GaAs) appear, namely, α-GaAs-LO mode peaking at ~240 cm$^{-1}$ (in purple) and α-GaAs-TO mode peaking at ~260 cm$^{-1}$ (in green)[58, 59]. Evidently, the α-GaAs phase appears only in the indented GaAs:Zn slice, indicating that the presence of a high enough concentration of holes is responsible for the facilitated amorphisation in the indented GaAs.

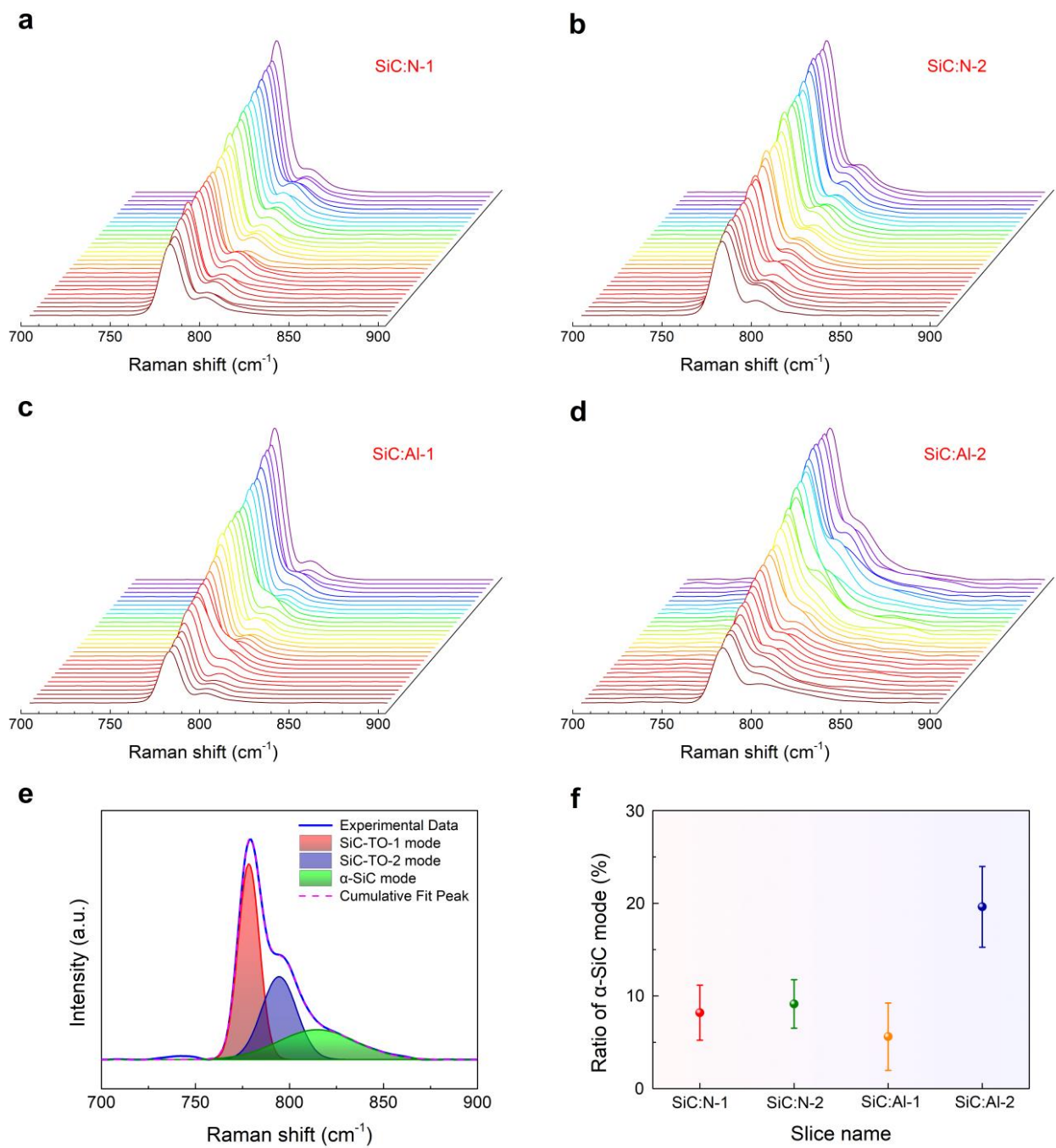

**Extended Data Fig. 4| The pressure-induced phase transformation behaviors in various 4H-SiC slices. a-d,** Raman spectra acquired from the 30 micro-indentations imposed on the lightly nitrogen (N)-doped SiC (SiC:N-1, n-type) slice with an electron concentration of $2.40 \times 10^9$ cm$^{-3}$, heavily N-doped SiC (SiC:N-2, n-type) slice with an electron concentration of $2.43 \times 10^{19}$ cm$^{-3}$, lightly aluminum (Al)-doped SiC (SiC:Al-1, p-type) slice with a hole concentration of $1.03 \times 10^{16}$ cm$^{-3}$, and heavily Al-doped SiC (SiC:Al-2, p-type) slice with a hole concentration of $1.75 \times 10^{19}$ cm$^{-3}$, respectively. **e,** Deconvolution of a representative Raman spectrum of the indented SiC, manifesting with a broad amorphous SiC (α-SiC) band peaking at ~815 cm$^{-1}$ (in green) and two relatively narrower bands corresponding to SiC-TO modes peaking at ~778 cm$^{-1}$ (in red) and at ~794 cm$^{-1}$ (in blue)[58, 60, 61],

respectively. **f,** Statistical results of the ratios of α-SiC mode in the Raman spectra shown in **a-d.** Herein, the ratio of α-SiC mode in a Raman spectrum is defined as the ratio of the integrated intensity of α-SiC band to that of the full spectrum. Note that the average ratio of α-SiC mode for SiC:Al-2 slice is distinctly larger than those for other three slices, indicating that the amorphisation in the indented SiC is facilitated by the presence of a high enough concentration of holes.

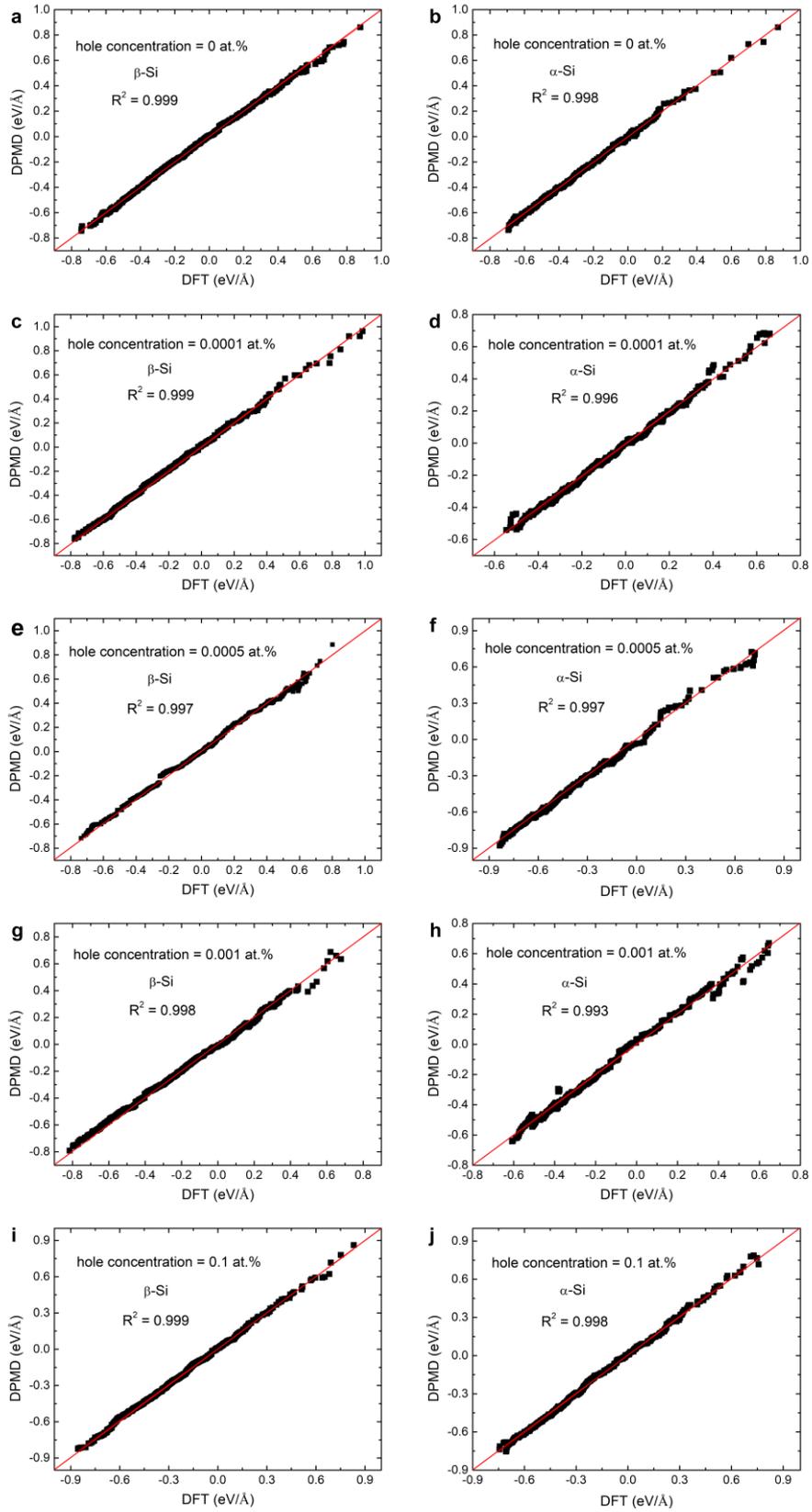

(Extended Data Fig. 5 -continued)

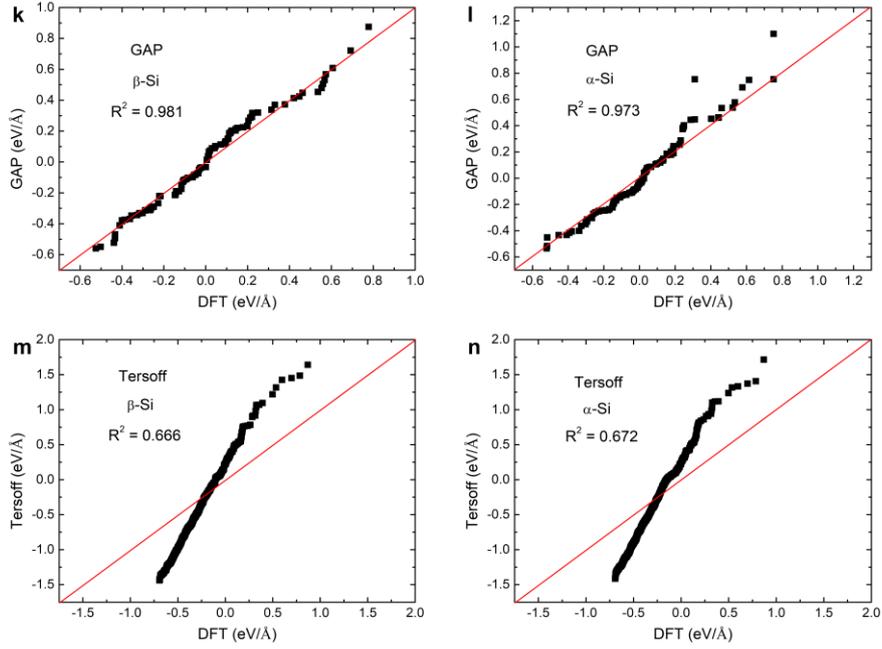

**Extended Data Fig. 5| Normalized changes in atomic forces for the Si-II (β-Si) and amorphous Si (α-Si) phases in different systems predicted by DeePMD potentials, Gaussian approximation potentials (GAP) or Tersoff potentials, compared with those derived from DFT calculations.** DeePMD potentials are applied to the systems with different hole concentrations. While, the GAP and Tersoff potentials are only applied to the non-doped systems.

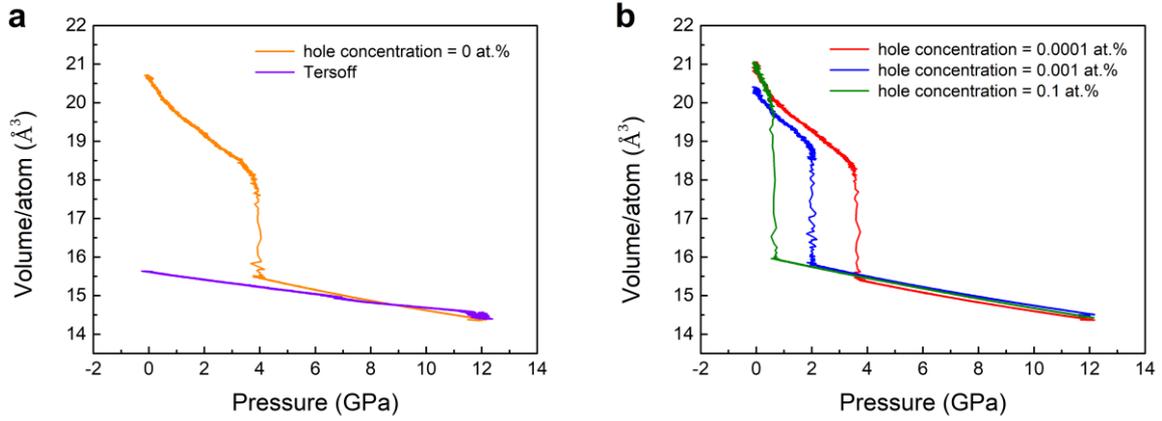

**Extended Data Fig. 6| Simulations on the changes in volume with pressure during the decompression process starting from Si-II phase for the systems with different hole concentrations. a,** The simulations with Tersoff potential (purple line) and DeePMD potential (orange line) for the system without holes (hole concentration = 0 at.%). Note that only DeePMD potential reproduces the structural phase transition. **b,** The simulations with the DeePMD potentials for the systems with different hole concentrations. Note that the abrupt change in volume as shown in **a** or **b** indicates the occurrence of phase transformation.

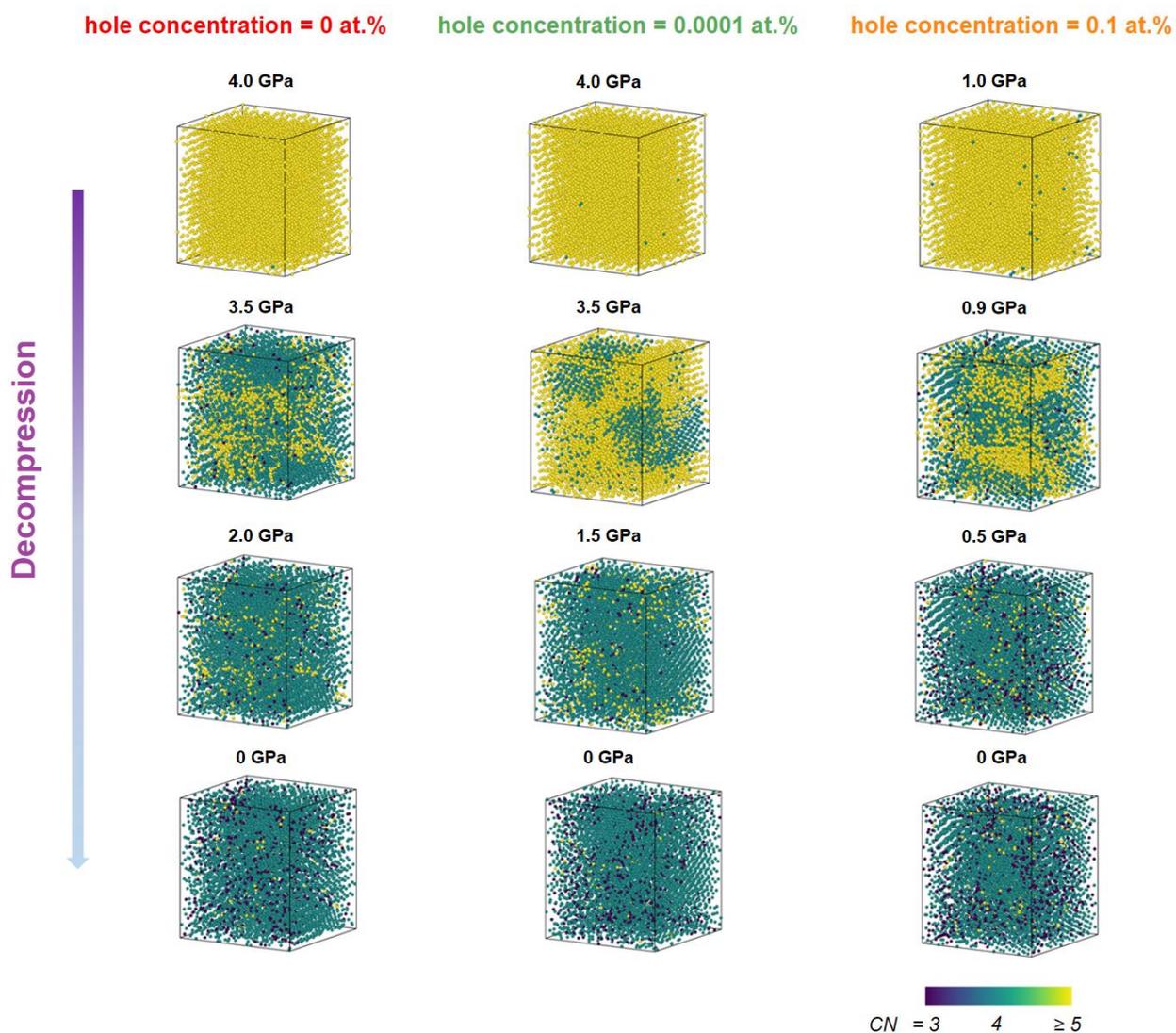

**Extended Data Fig. 7| Structural evolutions during the decompression process starting from the Si-II phase for the three systems with different hole concentrations.** The simulation cells are shown in 3D view, offering the same perspective in all views. Coordination numbers, CN (spatial cut-off = 3.0 Å), are indicated by the color coding. All structural drawings were created using OVITO.

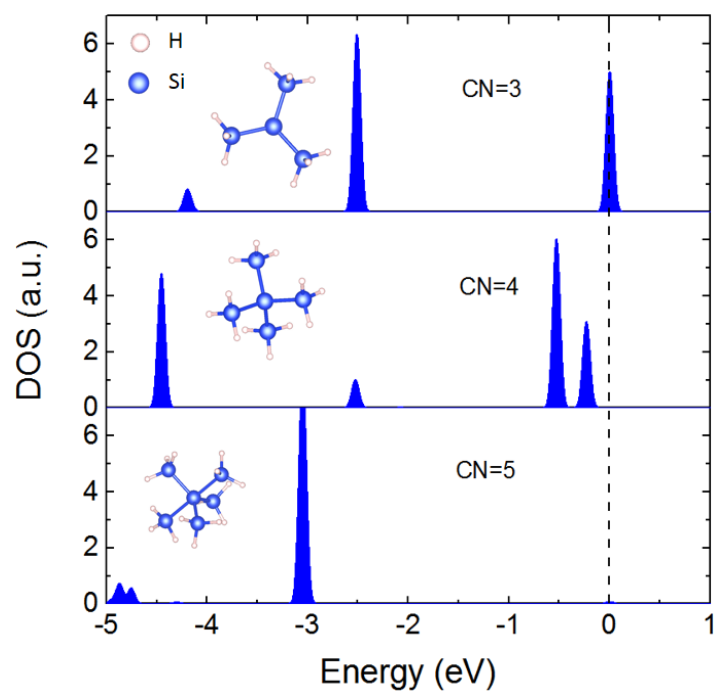

**Extended Data Fig. 8| Projected DOS of the Si atoms with different coordination numbers.** The three insets schematically illustrate the aperiodic Si structures in which the core Si atoms are bonded with three (CN = 3), four (CN = 4), or five (CN = 5) surrounding Si atoms, respectively. All of the surrounding Si atoms are passivated by three hydrogen atoms.

**Extended data Table 1| The <100>- oriented Si slices doped with boron (B) or phosphorus (P).**

| Dopant | Slice No. | Resistivity (Ω·cm) | Carrier concentration | |
|---|---|---|---|---|
| | | | cm$^{-3}$ | at.% |
| Boron (B) | 1 | $1.53 \times 10^4$ | $2.81 \times 10^{11}$ | $5.62 \times 10^{-10}$ |
| | 2 | 16.30 | $8.04 \times 10^{14}$ | $1.61 \times 10^{-6}$ |
| | 3 | 2.40 | $2.77 \times 10^{15}$ | $5.54 \times 10^{-6}$ |
| | 4 | 1.25 | $1.15 \times 10^{16}$ | $2.30 \times 10^{-5}$ |
| | 5 | 0.115 | $1.99 \times 10^{17}$ | $3.98 \times 10^{-4}$ |
| | 6 | 0.075 | $3.59 \times 10^{17}$ | $7.18 \times 10^{-4}$ |
| | 7 | 0.055 | $5.63 \times 10^{17}$ | $1.13 \times 10^{-3}$ |
| | 8 | 0.044 | $7.85 \times 10^{17}$ | $1.57 \times 10^{-3}$ |
| | 9 | 0.024 | $2.00 \times 10^{18}$ | $4.00 \times 10^{-3}$ |
| | 10 | 0.015 | $4.07 \times 10^{18}$ | $8.14 \times 10^{-3}$ |
| | 11 | 0.0076 | $1.19 \times 10^{19}$ | $2.38 \times 10^{-2}$ |
| | 12 | 0.0028 | $4.05 \times 10^{19}$ | $8.10 \times 10^{-2}$ |
| Phosphorus (P) | 13 | 0.0011 | $6.52 \times 10^{19}$ | 0.13 |

Note: The resistivity of a Si slice was measured by four-point probe. The corresponding carrier concentration was transformed from the resistivity according to ASTM F723-88.

**Extended data Table 2| The <100>-oriented Si slices codoped with boron (B) and germanium (Ge).**

| Slice No. | Resistivity (Ω·cm) | [B] (cm$^{-3}$) | [Ge] (cm$^{-3}$) |
|---|---|---|---|
| Si:(B,Ge)-1 | 0.0072 | $1.28 \times 10^{19}$ | $5.52 \times 10^{19}$ |
| Si:(B,Ge)-2 | 0.0066 | $1.45 \times 10^{19}$ | $7.54 \times 10^{19}$ |
| Si:(B,Ge)-3 | 0.0062 | $1.57 \times 10^{19}$ | $8.25 \times 10^{19}$ |

Note: The Si codoped with B and Ge impurities is denoted as Si:(B,Ge). The resistivity of a Si slice was measured by four-point probe. The B concentration ([B]) was approximately equal to the carrier concentration transformed from the resistivity according to ASTM F723-88. The Ge concentration ([Ge]) was measured by secondary ion mass spectroscopy (SIMS).

**Extended data Table 3| The <100>-oriented Si slices codoped with B and P.**

| Slice No. | Resistivity (Ω·cm) | [B] (cm$^{-3}$) | [P] (cm$^{-3}$) |
|---|---|---|---|
| Si:(B,P)-1 | 0.46 | $1.28 \times 10^{19}$ | $1.23 \times 10^{19}$ |
| Si:(B,P)-2 | 0.0039 | $1.35 \times 10^{19}$ | $3.69 \times 10^{19}$ |

Note: The Si codoped with B and P impurities is denoted as Si:(B,P). The concentrations of B and P impurities ([B] and [P]) were measured by SIMS. The Si:(B,P)-1 is of p-type conduction and the Si:(B,P)-2 is of n-type conduction, determined by the thermo-probe method. The resistivity of a Si slice was measured by four-point probe.

**Extended data Table 4| The <100>-oriented Si slices doped with gallium (Ga).**

| Slice No. | Resistivity (Ω·cm) | Carrier concentration (cm$^{-3}$) |
|---|---|---|
| Si:Ga-1 | 0.45 | $3.61 \times 10^{16}$ |
| Si:Ga-2 | 0.027 | $1.67 \times 10^{18}$ |

Note: The Si doped with Ga impurities is denoted as Si:Ga. The resistivity of a Si slice was measured by four-point probe and the corresponding hole concentration was transformed according to ASTM F723-88.

**Extended data Table 5| The <100>-oriented near-intrinsic and Ga-doped Ge slices.**

| Conduction type | Slice No. | Resistivity (Ω·cm) | Carrier concentration (cm$^{-3}$) |
|---|---|---|---|
| Near-intrinsic | NI-Ge | 50.0 | $2.93 \times 10^{13}$ |
| p-type | Ge:Ga | 0.0066 | $2.08 \times 10^{18}$ |

Note: The near-intrinsic and Ga-doped Ge are denoted as NI-Ge and Ge:Ga, respectively. The resistivity and carrier concentration of a Ge slice were obtained by Hall effect measurement.

**Extended data Table 6| The <100>-oriented semi-insulating, p-type and n-type GaAs slices.**

| Conduction type | Slice No. | Resistivity (Ω·cm) | Carrier concentration (cm$^{-3}$) |
|---|---|---|---|
| semi-insulating type | SI-GaAs | / | / |
| p-type | GaAs:Zn | 0.0059 | $1.48 \times 10^{19}$ |
| n-type | GaAs:Si | 0.0015 | $1.31 \times 10^{18}$ |

Note: The semi-insulating GaAs, zinc-doped, or Si-doped GaAs is denoted as SI-GaAs, GaAs:Zn or GaAs:Si.

The resistivity and carrier concentration of a GaAs slice were obtained by Hall effect measurement.

**Extended data Table 7| The <0001>-oriented p- and n-type SiC slices.**

| Conduction type | Slice No. | Resistivity (Ω·cm) | Carrier concentration (cm$^{-3}$) |
|---|---|---|---|
| n-type | SiC:N-1 | $4.35 \times 10^6$ | $2.40 \times 10^9$ |
|  | SiC:N-2 | 0.015 | $2.43 \times 10^{19}$ |
| p-type | SiC:Al-1 | 33.18 | $1.03 \times 10^{16}$ |
|  | SiC:Al-2 | 0.181 | $1.75 \times 10^{19}$ |

Note: The nitrogen- or aluminum-doped SiC is denoted as SiC:N or SiC:Al. The resistivity and carrier concentration of a SiC slice were obtained by Hall effect measurement.

**Extended data Table 8| Root-mean-square error in energies (RMSEE) and forces (RMSEF) predicted by five DeePMD models with respect to the DFT calculations for training and testing datasets.**

| hole concentration (at.%) | $RMSE_E$, meV/Å | | $RMSE_F$, meV/atom | |
| --- | --- | --- | --- | --- |
| | Train | Test | Train | Test |
| 0 | 6.12 | 3.35 | 79.9 | 86.3 |
| 0.0001 | 1.23 | 5.18 | 58.5 | 114 |
| 0.0005 | 3.71 | 2.29 | 56.3 | 141 |
| 0.01 | 1.51 | 3.94 | 66.5 | 147 |
| 0.1 | 1.82 | 2.31 | 133 | 144 |

Note: The RMSE of each training set is obtained after stabilization with numbers of training steps.